# High-fidelity, low-latency polarization quantum state transmissions over a hollow-core conjoined-tube fibre at around 800 nm

Xin-Yu Chen,[1,4,†] Wei Ding,[2,†,*] Ying-Ying Wang,[2,3] Shou-Fei Gao,[2] Fei-Xiang Xu,[1,4] Hui-Chao Xu,[1,4] Yi-Feng Hong,[3] Yi-Zhi Sun,[2] Pu Wang,[3] Yan-Qing Lu,[1,4,*] and Lijian Zhang[1,4,*]

[1]*National Laboratory of Solid State Microstructures and College of Engineering and Applied Sciences, Nanjing University, Nanjing 210093, China*
[2]*Institute of Photonics Technology, Jinan University, Guangzhou 510632, China*
[3]*Institute of Laser Engineering, Beijing University of Technology, Beijing 100124, China*
[4]*Collaborative Innovation Center of Advanced Microstructures, Nanjing University, Nanjing 210093, China*
[†]These authors contributed equally.
* photonicsweiding@163.com, * lijian.zhang@nju.edu.cn, * yqlu@nju.edu.cn

**Abstract:** The performances of optical fibre-based quantum information systems are limited by the intrinsic properties of silica glass materials, e.g. high latency, Rayleigh-scattering loss wavelength scaling law, and cross-coupling induced modal impurity. Hollow-core optical fibre (HCF) promises to unify air-borne light propagation and non-line-of-sight transmission, thus holding great potentials for versatile photonics-based quantum information applications. The early version of HCF based on photonic-bandgap guidance has not proven itself as a reliable quantum channel because of the poor modal purity in both spatial and polarization domains, as well as significant difficulty in fabrication when the wavelength shifts to the visible region. In this work, based on the polarization degree of freedom, we first, to the best of our knowledge, demonstrate high-fidelity (~0.98) single-photon transmission and distribution of entangled photons over a conjoined-tube hollow-core fibre (CTF) by using commercial silicon single-photon avalanche photodiodes. Our CTF realized the combined merits of low loss, high spatial mode purity, low polarization degradation, and low chromatic dispersion. We also demonstrate single-photon low latency (~99.96% speed of light in vacuum) transmission, thus paving the way for extensive uses of HCF links in versatile polarization-based quantum information processing.

## 1. Introduction

Solid-core silica glass fibres have been widely used as the transmission medium in quantum information systems[1–3]. As the distance of quantum communication[4,5] continuously elongates and the number of quantum nodes[6] steadily increases, we are now at an exciting moment to await the birth of multi-functional quantum internets[7] which are anticipated to surmount classical counterparts in a number of aspects. With such a prospect, the intrinsic limitation imposed by the material of optical fibres will ultimately become a fundamental obstacle to the performance upgrade of current quantum link technology.

Compared with free-space links, the reduced speed of light in a fibre link may impair loophole-free Bell inequality test[8] and device-independent quantum randomness generation[9] between remote notes, and add the time overhead of quantum communications. In a variety of time-and-synchronization-sensitive applications such as distributed quantum computation and quantum repeaters, the fibre latency will retard the formation of entanglement and decrease the efficiency of quantum links.

Secondly, the low-loss transmission window of silica glass fibres is jointly determined by the Rayleigh scattering loss (RSL) and the multi-phonon absorption. While the photon-atom interactions used in quantum memory and quantum information processing[10,11] mostly occur in the visible region, and the mature and cost-effective silicon single-photon avalanche photodiodes (Si-SPADs)[12] offers high detection efficiencies, high count rates and low noise in the wavelength range of 400-1000 nm, the $\lambda^{-4}$ dependence of the RSL renders the working wavelengths of photon transmission, manipulation and detection incompatible with each other, therefore necessitating a sophisticated photon conversion step[13].

Last but not least, the fact that the total internal reflection (TIR) light guidance in a solid-core fibre relies on a tight overlap of optical field with glass brings about a lot of detrimental effects, such as inter-modal coupling, polarization degradation and chromatic dispersion (CD), resulting in serious decoherence in fibre-based quantum information systems.

Fortunately, light guidance can also be realized in a void or gaseous fibre channel with the aid of photonic micro-structures[14]. Liberated from the TIR mechanism, hollow-core optical fibres (HCFs) outperform solid-core fibres in terms of much reduced latency and mode field overlap with glass, which provide a novel means to solve the problems mentioned above and to enable light waves immune from external fluctuations caused by the thermo-optic and elasto-optic effects in solid glass[15].

The first generation of HCF relied on the formation of photonic bandgap (PBG)[16]. The dominant attenuation in PBG-HCF comes from the surface scattering loss (SSL)[17] and sticks to the level of around 1.2 dB km$^{-1}$ at telecom wavelengths. Moreover, the poor modal purity[18] and the challenging fabrication accompanying with geometry downscaling[19] both prohibit the employment of PBG-HCF in quantum information systems.

In contrast, another class of HCFs based on anti-resonant (AR) reflecting optical waveguide guidance[20] (or inhibited coupling[21]) recently experience a rapid performance improvement[22], thanks to the introductions of negative-curvature core boundary[23] and multiple cladding layers[24,25]. In 2018, two geometries of low-loss AR-HCF emerged with the conjoined-tube fibre (CTF) exhibiting the minimum loss of 2 dB km$^{-1}$ at 1512 nm[26] and the nested AR nodeless fibre (NANF) realizing the minimum loss of 1.3 dB km$^{-1}$ at 1450 nm[27]. The latest loss record of NANF has reached 0.28 dB km$^{-1}$ across the C and L bands[28], considerably narrowing the gap with solid-core glass fibres to a factor less than two. Furthermore, the single-spatial-mode attribute of AR-HCF has also been validated in experiments[29,30], and an exceptional low inter-polarization coupling (orders of magnitude lower than solid-core glass fibres) is discovered recently[31].

Besides the superiority in loss and modal purity, in comparison with PBG-HCF, AR-HCF also features convenient fabrication with geometry downscaling and a $\lambda^{-1}$ dependence of the minimum loss, both of which facilitate its use at shorter wavelengths. Recently, these two features have been demonstrated in two visible-guiding CTFs with the RSL of silica glass fibre being surpassed[32]. By contrast, the RSL in solid-core glass fibres and the SSL in PBG-HCFs are fundamentally hampered by the wavelength scaling laws of $\lambda^{-4}$ and $\lambda^{-3}$, respectively.

In this work, we design and fabricate a low-loss, 800 nm-guiding CTF, which simultaneously possesses high mode purity, low polarization degradation, low CD, and low latency. In this novel photonic quantum link free from frequency conversion, great resistance to decoherence is demonstrated by high-fidelity single-photon polarization state transmission and polarization entanglement distribution over a 36.4 m length. Our CTF link exhibits superb quality comparable with a conventional solid-core fibre link. Moreover, a single-photon time-of-flight measurement also confirms the low-latency attribute[33] of our CTF as quantum links.

## 2. Results

### 2.1 Fabricated CTF

As displayed in Fig. 1a, the CTF consists of a 29.3 μm diameter hollow core, six non-touching conjoined tubes (CTs), and a 230 μm diameter outer sheath. High-resolution scanning electron microscope (SEM) images of the fibre cross-section reveal that the thicknesses of the glass membranes are $t_{1,2,3} \approx 606 \pm 26 / 507 \pm 26 / 818 \pm 35$ nm, and the areas of the cladding air holes are $A_{1,2} \approx 196.6 \pm 20 / 66.7 \pm 8.7$ μm$^2$. According to the AR reflecting optical waveguide model[20] and the coupled-mode theory[34], the two layers of glass membranes close to the core (i.e. the negative-curvature core boundary one and the one inside the CTs) and the two layers of air holes in the cladding can provide coherent back-reflections[25,35] at wavelengths around 800 nm, where our CTF works in the second-order AR condition. The azimuthal gaps between adjacent CTs are measured to be 6.05 ± 1.15 μm (~21% of the core diameter), meaning that light leakage through these gaps is modest[36]. Figure 1b shows a spectral loss measured by the cutback method (see Methods) with a transmission window ranging from 750 nm to 900 nm and the minimum loss of 9.2 ± 0.7 dB km$^{-1}$ at 845 nm.

To achieve high modal purity in both spatial and polarization domains, low environmental sensitivity, and low chromatic dispersion, we incorporated the following three features into the geometry of our CTF. Firstly, a big core diameter to wavelength ratio (~35) was adopted to guarantee a less than 10$^{-4}$ mode field overlap with glass. Secondly, the flat glass membranes inside the CTs were arranged away from the hollow core as far as possible (see Methods), rendering the polarization purity of the core mode not to be severely affected by Fano resonances existing around these structures[37]. Very lately, a remarkable feature of weak inter-polarization coupling (orders of magnitude weaker than the fundamental anisotropic Rayleigh scattering limit in glass core fibres[38]) were found in statically-placed AR-HCFs[31]. This finding opened a new prospect for exploiting polarization resources in light transmission along AR-HCF. For our CTF geometry, simulation based on finite element method (FEM) revealed that the phase birefringence and the polarization-dependent loss (PDL) were in the levels of 10$^{-7}$ and <5 dB km$^{-1}$, respectively, see Fig. 1c. Additionally, in order to attain a high spatial mode purity in our CTF, an optimized light in-coupling condition and the resonant filtering of higher-order modes induced by the first ring of cladding holes had been employed. Thirdly, compared with our previously reported visible CTFs[32], the glass membrane thicknesses of this CTF ($t_1$ and $t_2$) support a lower-order AR band at 800 nm, therefore yielding a wider transmission window and a lower CD.

### 2.2 Single-photon polarization state transmission

Having prepared a low loss, high spatial mode purity, and low polarization degradation link in our CTF, we built a heralded single-photon source at 830 nm based on spontaneous parametric down conversion (SPDC) with the detected rate of about 30,000 counts per second (cps) after the collecting single-mode fibres (SMFs) and the heralding efficiency of 16% (see Methods). By arranging a linear polarizer and two wave plates (half-wave plate, HWP, and quarter-wave plate, QWP), we projected the single photons onto the six polarization states (H/V/D/A/R/L, see Methods for the definitions) to form a complete set of mutually unbiased bases and then injected them into the CTF (see Fig. 1d) with the overall transmittance of above 60% (including in-coupling/out-coupling efficiencies and propagation loss along the fibre). The polarization rotation accumulated in the transmission via the fibre and the optics was then compensated by the wave plates.

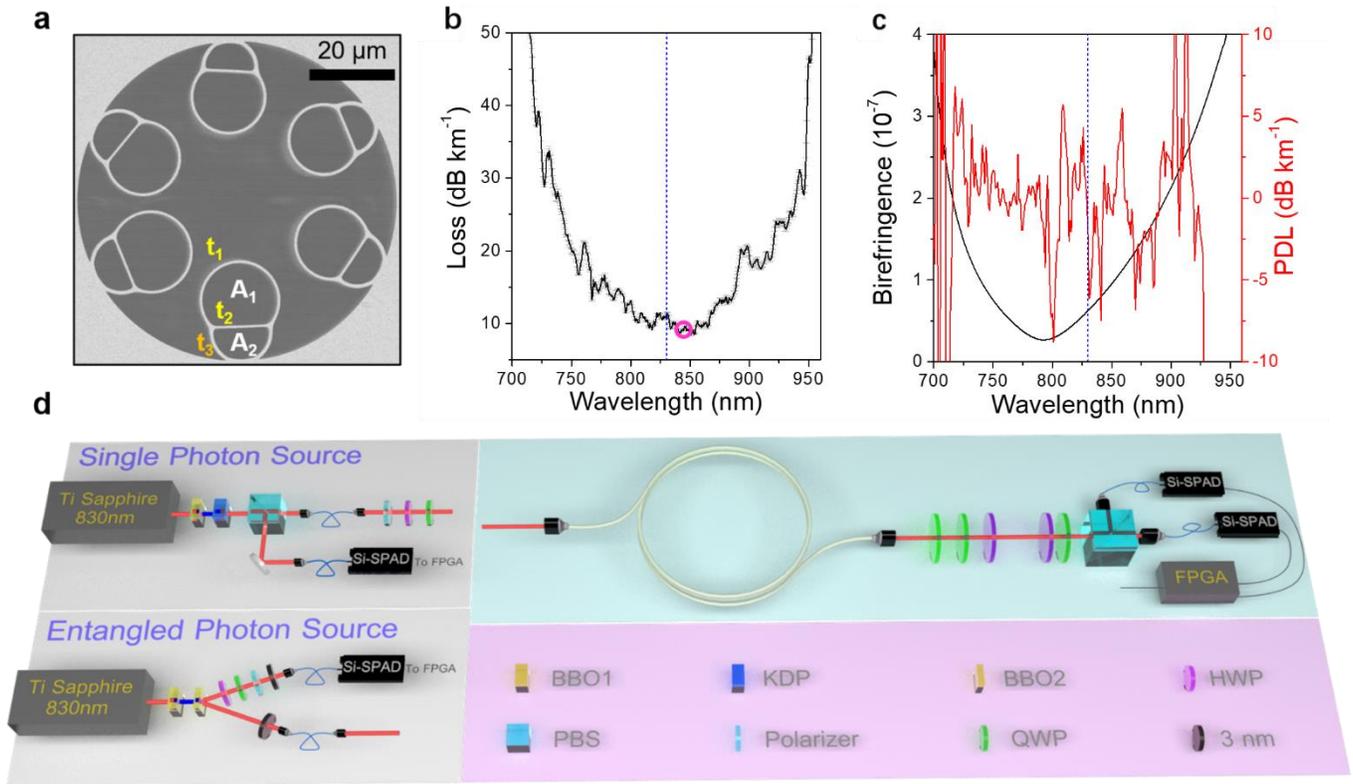

**Figure 1 Characterization of the CTF and experimental setup for polarization quantum state transmission. a,** SEM image of the CTF. **b,** Cutback measured loss spectrum. The minimum loss and the quantum transmission wavelength are marked by the purple circle and the blue dashed line, respectively. **c,** Simulated phase birefringence and PDL based on cross-section micrographs. **d,** Experimental setup consisting of a heralded single-photon source (upper left), a polarization entangled photon-pair source (lower left) and a fibre-testing bench. BBO1/BBO2, β-barium-borate crystals for frequency doubling and generation of entangled photon-pairs, respectively; KDP, potassium dihydrogen phosphate crystal for generation of single photons; HWP, half-wave plate; QWP, quarter-wave plate; PBS, polarization beam splitter with the extinction ratio of over 30 dB; Polarizer, linear polarizer with the extinction ratio of over 40 dB.

We also reconstructed the transmitted polarization states in the context of quantum state tomography[39] (QST) by using two Si-SPADs, whose detection efficiencies and dark count rates at 830 nm were over 60% and less than 150 cps, respectively (see Methods). In our experiments, the single-photon count rates after transmission were ~10,000 cps, the average fidelity between the transmitted and the input states was ~0.980, and the average purity was ~0.998, as shown in Fig. 2a. It was worth noting that all the fidelities secured in our measures were far above the classical limit of 2/3. As a control experiment, we also conducted QST for the same sets of polarization states through a piece of SMF (Corning, HI780) with the same length. The average fidelity and purity of the transmitted states were about 0.975 and 0.997, respectively. Both these two fibres under test were statically and loosely placed on a table with a radius of 25 cm. The results revealed that our CTF had the same level of polarization preserving capability as the SMF, thanks to the excellent spatial and polarization mode purities in a low mode-field-glass overlapping fibre.

To further test our CTF link, quantum process tomography (QPT)[40] was implemented to reconstruct the χ matrix of the transmitted photons with $\varepsilon(\rho) = \sum_{m,n=0}^{d^2-1} \chi_{mn} \hat{E}_m \rho \hat{E}_n^\dagger$. Here, $\{\hat{E}_i\}$ represent the Pauli bases in one-qubit cases, and ρ is the density matrix of the input states. As shown in Fig. 2b, the QPT measurement yields a fidelity of 0.983 with respect to the identity process. In comparison, the QPT for the SMF link gives out a fidelity of 0.972.

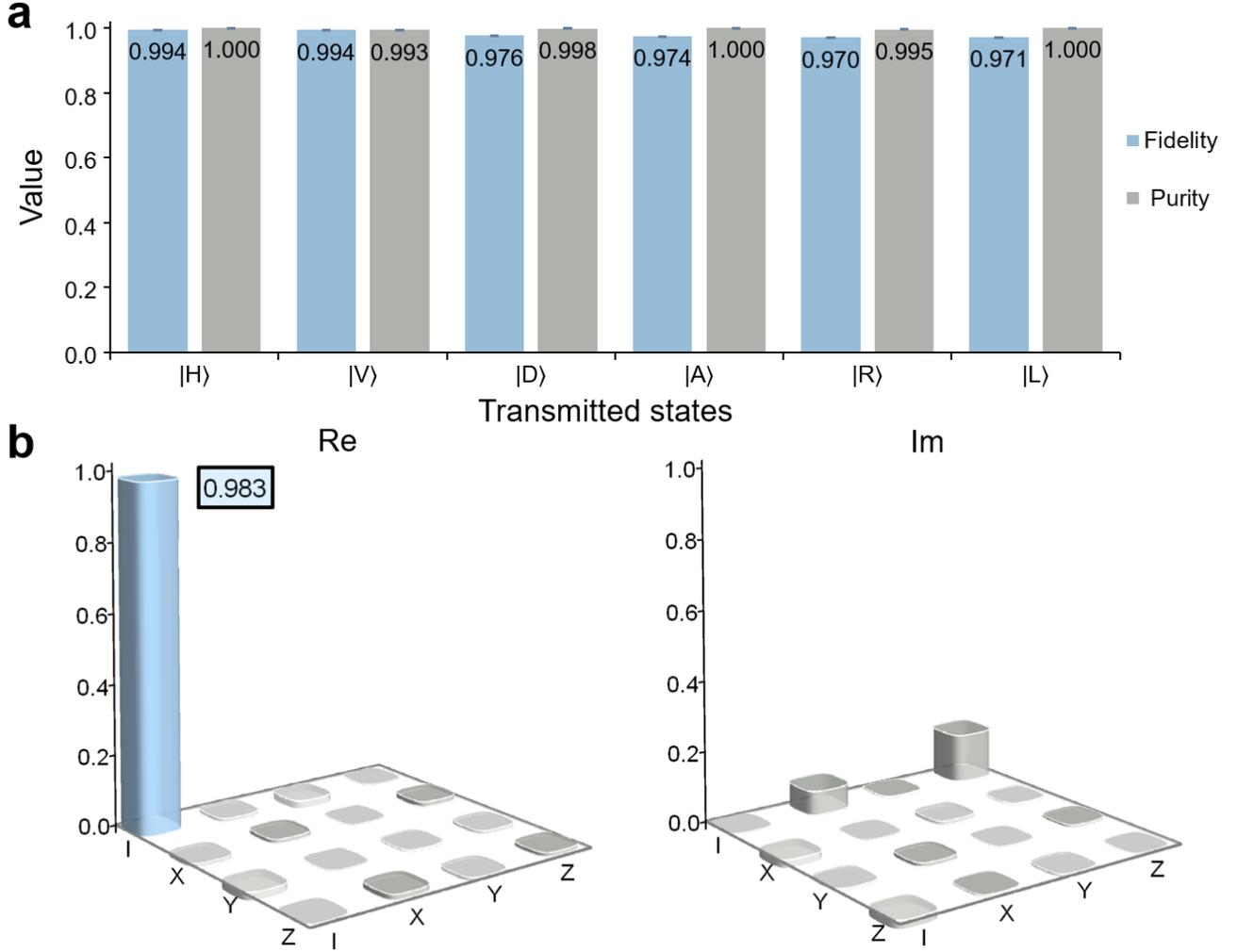

**Figure 2 Experimental results of single-photon polarization state transmission. a,** Fidelities and purities of the six transmitted polarization states. The average fidelity and purity are ~0.980 and ~0.998, respectively. The error bars represent one standard deviation. **b,** Process matrix of the 36.4 m CTF link, revealing a fidelity of 0.983 with respect to an identity process matrix.

### 2.3 Polarization entanglement distribution

To demonstrate distribution of polarization entanglement over the CTF link, an entangled photon-pair source with the singlet state of $|\psi^-\rangle = \frac{1}{\sqrt{2}}(|H\rangle_1|V\rangle_2 - |V\rangle_1|H\rangle_2)$ and the fidelity of 0.956 had been constructed also at 830 nm via type-II SPDC (see Fig. 1d). One photon was measured locally, and the other was transmitted through the fibre link. The coincidence counts between the two Si-SPADs in the local node and at the output end of the tested fibre reached to ~10,000 per 10 seconds.

The polarization correlations between the two photons were measured by projecting the local photon onto H, V, D, A states and rotating the polarization angle of the transmitted photon. As shown in Fig. 3a, the visibilities of the polarization correlations after the CTF link (without subtracting the background noise) were 97.1% and 97.8% in the H/V and the D/A bases, respectively, well above the threshold value of 71% for the Bell inequality violation. We also reconstructed the entangled state with the density matrix shown in Fig. 3b. The fidelity between the distributed entangled state and the initial state was 0.977. To highlight the quality of the distributed entanglement, we evaluated the Clauser-Horne-Shimony-Holt (CHSH) Bell's inequality by the S parameter[41], whose result of 2.741 ±0.018 violated the classical limit of 2 by 41 standard deviations. All the above results confirmed that polarization entanglement had been maintained very well across the full length of CTF. In combination with the above single-photon transmission tests, our experiments unambiguously prove that our CTF can be used as a reliable link for transmission of quantum information encoded in the polarization degree of freedom of photons.

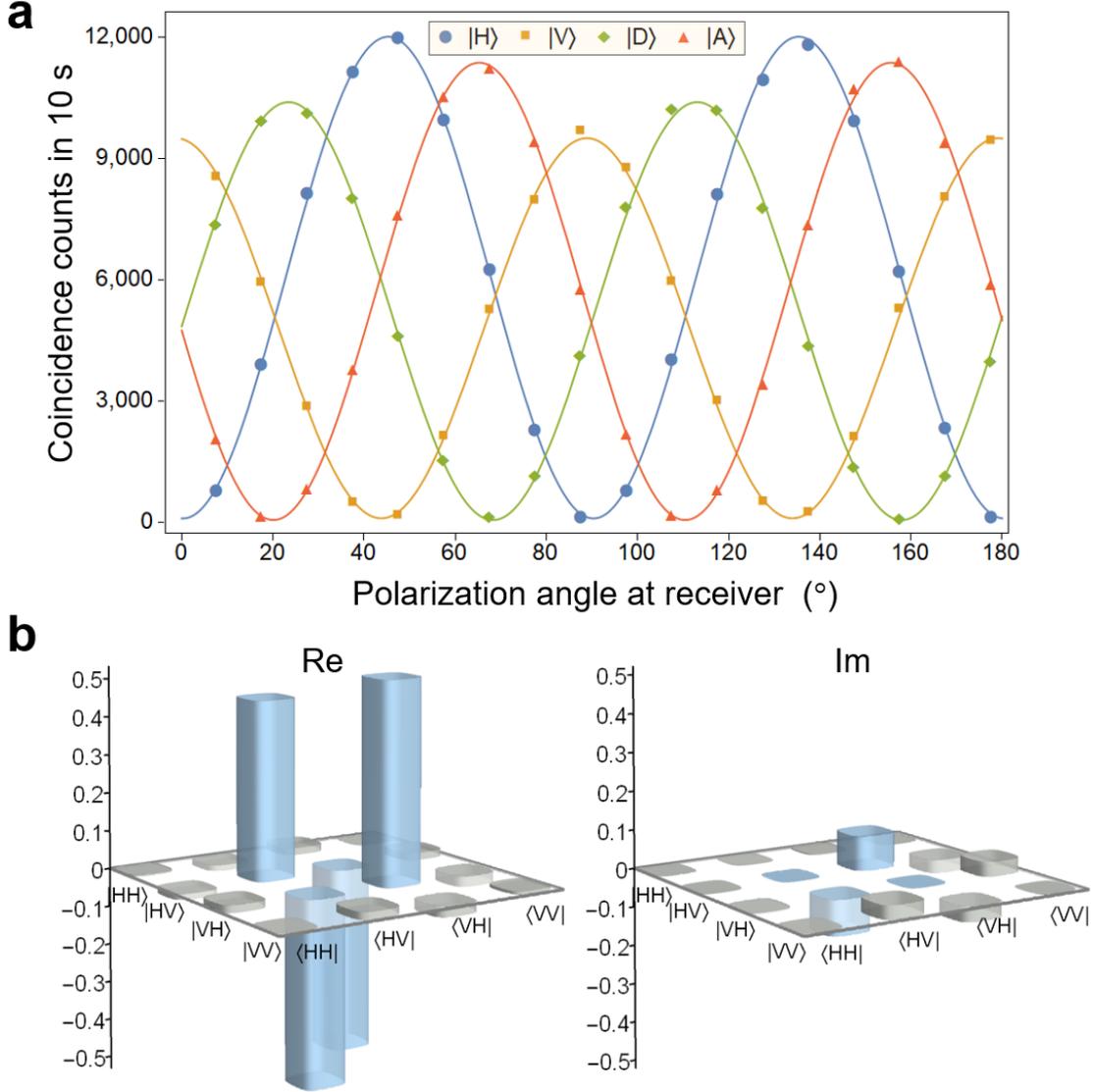

**Figure 3 Experimental results of polarization entanglement distribution. a,** Polarization correlations between the local and the transmitted photons. The visibilities in the H/V and the D/A bases are measured to be 97.1% and 97.8%, respectively. **b,** Density matrix of the distributed entangled state. The fidelity with respect to the initial state is 0.977.

### 2.4 Chromatic dispersion and latency

To demonstrate the intrinsic merit of HCF, we measured the group velocity dispersion and the group index of a short piece of CTF using low-coherence interferometry. As shown in Fig. 4a, at 830 nm, our CTF had a fifty times lower dispersion than the SMF and almost the same (99.96%) speed of light as in vacuum. Such a low dispersion should be attributed to the low mode field overlap with glass and the wide transmission window[30,42] of the second-order AR guidance. The big core diameter relative to the wavelength (with the ratio of ~35) explains the reason of the faster light speed in our CTF than in the reported 19-cell PBG-HCF[33] (with the ratio of ~17). By contrast, the high dispersion and high latency in the control SMF are fundamentally caused by the material properties of silica glass and the TIR guidance.

To manifest the low-latency feature of our CTF link, a single-photon time-of-flight measurement was implemented. As shown in Fig. 4b, over a fibre distance of $L_{CTF} \approx L_{SMF} \approx 36.4 \pm 0.1$ m, a time leading of $(n_g^{(SMF)} \cdot L_{SMF} - n_g^{(CTF)} \cdot L_{CTF})/c = 55.4 \pm 0.6$ ns was accumulated in our CTF link, yielding a group index difference of $n_g^{(SMF)} - n_g^{(CTF)} = 0.456 \pm 0.009$. Here, $c$ is the speed of light in vacuum, $n_g^{(CTF)} \approx 1$ is hypothesized, and 0.6 ns represents the uncertainty from the timing jitter in the two Si-SPADs. It is worth noting that the two measurements of group index difference in Fig. 4a and Fig. 4b agree well with each other.

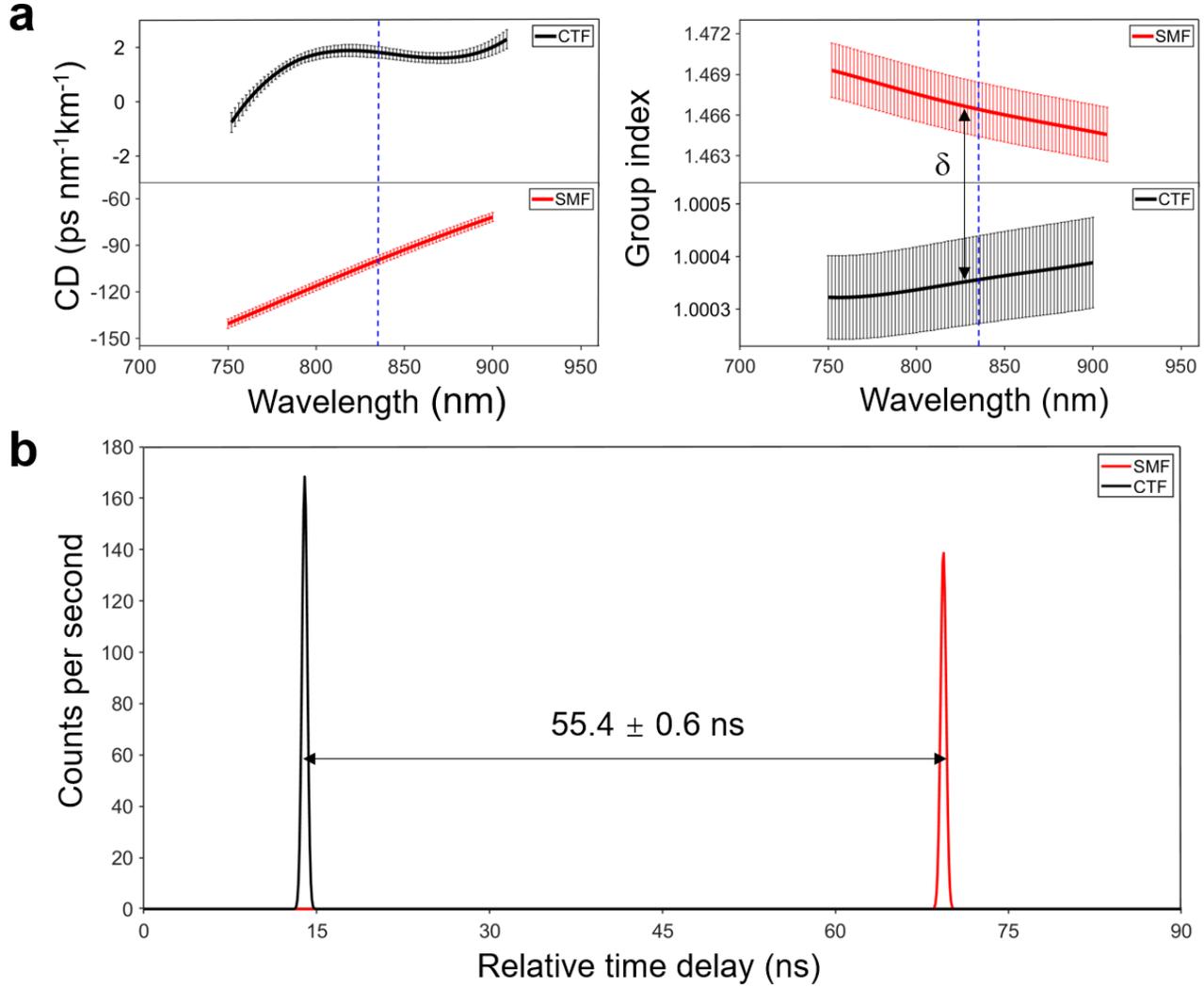

**Figure 4 Characterization of chromatic dispersion, group index, and single-photon latency. a,** Measured group velocity dispersions and group indices of a 20 cm-length CTF and a 5 cm-length SMF using a free-space Mach–Zehnder interferometer. δ = 0.466 ±0.003. **b,** Relative time delay of single photons transmitting through the CTF and SMF with the same length of 36.4 ±0.1 m. The timing jitter induced uncertainty is estimated to be 0.6 ns.

## 3. Discussion

Quantum internet is one of the ultimate targets of quantum information science and technology. For realizing this goal, reliable generation, transmission, manipulation, and detection of quantum states are the keystones. We deem that the high-fidelity single-photon polarization transmission and polarization entanglement distribution through a CTF at Si-SPAD wavelengths as demonstrated in this work are just an early indication of more versatile use of advanced AR-HCFs in quantum information processing with more degrees of freedom of photons. Apart from permitting qubits to fly with the fastest speed in the universe, AR-HCFs also provide an unprecedented way to exploit the resource of optical fibres associated with spatial/polarization mode purities and transmission window in the visible region, both facilitating quantum information encoding in photons.

To realize the above functions, AR-HCFs have to be designed and fabricated deliberately. As shown in this proof-of-principle work, the flexibility provided by the cladding micro-structures in our CTF allows for simultaneously realizing low loss, high mode purity, low polarization degradation, and wide transmission window (so that reduced dispersion) at 800 nm. Over a distance of 36.4 m, we had examined the CTF-based polarization quantum state transmission and had not observed any inferiority compared with conventional fibre links.

In summary, while the TIR guiding mechanism discovered by John Tyndall in 1870 liberated light propagation from the line-of-sight way, and the high transparency of purified silica glass discovered by Charles Kao in 1966 opened long-distance fibre-optic applications, the development of HCF, which was triggered by Philip Russell's idea of incorporating micro-structures into the cladding in 1991, led us to another leap of using optical fibres in a free-space-like but non-line-of-sight manner. Such a liberation is of crucial importance for harnessing photons as quantum information couriers.

**Appendix**

**A.1 CTF fabrication and loss measurement**

A slim glass slice was inserted into a Hereaus Suprasil F300 synthetic silica glass tube and drawn into capillaries. The width of the glass slice (19.5 mm) was slightly smaller than the inner diameter of the glass tube (20 mm) to ensure an off-center geometry. The CT capillaries were piled up into a stack and then drawn into canes and fibre. Using a trial-and-error procedure, we realized a CTF with the following features: 1) almost touching CTs with modest gaps (~21% of the core diameter) in between; 2) $A_1 > A_2$ so that the glass bars inside the CTs being far away from the core; and 3) appropriate glass membrane thicknesses for light guidance at 830 nm at the second AR order.

To measure the spectral loss, we cut the CTF from 60 m to 10 m with the input end butt-coupled with a super-continuum laser source (YSL Photonics, SC-5). The tested fibre was loosely looped with a radius of 30 cm, a CCD camera monitored the output mode profile, and an optical spectral analyzer (Ando, AQ6317) acquired the transmission spectra. Multiple cleavages at the output end showed little variations in the recorded spectra for both the 60 m and 10 m fibre lengths.

After loss measurement, we made several fibre cleavages and SEM inspections to select out a 36.4 m long CTF sample, whose size variations at both ends are less than 3%.

**A.2 Single-photon and entangled photon sources**

A heralded single-photon source and an entangled photon-pair source were produced by SPDCs in a potassium dihydrogen phosphate (KDP) crystal and a β-barium-borate (BBO2) crystal, respectively. The pump light of the SPDCs was generated in a BBO1 frequency doubling crystal driven by a Ti : Sapphire pulsed laser (830 nm, 76 MHz). The down converted photons were spectrally filtered by 3 nm band-pass filters and spatially filtered by single-mode optical fibres. The wavelengths of our photon sources matched well with the transmission window of our CTF, and the 2nd order correlation functions were examined to be $g^{(2)}(0) = 0.0397$, illustrating a very good single-photon attribute. Moreover, the Si-SPADs used in our experiments offered a nominal detection efficiency of over 60%, a dark count rate of less than 150 cps, and random timing jitter of around 400 ps.

**A.3 QST and QPT**

To quantify the density matrix of the transmitted polarization states, we implemented QST via a sequence of projective measurements. The six universal polarization states of $|H\rangle$, $|V\rangle$, $|D/A\rangle = (|H\rangle \pm |V\rangle)/\sqrt{2}$, and $|R/L\rangle = (|H\rangle \pm i|V\rangle)/\sqrt{2}$ with H/V representing the horizontal/vertical polarizations, were prepared by utilizing a linear polarizer (extinction ratio of over 40 dB), an HWP, and a QWP. At the output end of the CTF, after compensating for the polarization rotations, the single photons were projected onto three orthogonal bases [H/V, D/A, and R/L, i.e. a mutually-unbiased-bases (MUB)] by utilizing another set of HWP, QWP and PBS (extinction ratio of over 30 dB). Two Si-SPADs recorded the coincidence counts, which were then analyzed by a field-programmable gate array (FPGA) (gating of 4 ns). In order to reduce the measurement error, all the coincidence experiments were undertaken twice by exchanging the two Si-SPADs. Then, all the experimental data were analyzed by minimizing the sum of $\sum [p_j^{(k)} - Tr(\rho^{(k)} M_j)]^2$ subject to $Tr\{\rho^{(k)}\} = 1$ and $\rho^{(k)} \geq 0$. QST of the distributed entangled photon states was also implemented by applying two-qubit MUB measurements. The purity was defined as $Tr\{\rho^2\}$, and the fidelity between the experimental state ($\rho_e$) and the theoretical state ($\rho_t$) was calculated by $F_s(\rho_e, \rho_t) = (Tr\{\sqrt{\sqrt{\rho_e}\rho_t\sqrt{\rho_e}}\})^2$.

Based on the QST results, standard QPT was also conducted by linear inversing the measured data to achieve the process matrix $\chi$. For one qubit case, the identity process matrix denoted by $\chi_i$ should be a 4×4 matrix with $\chi_{11} = 1$ and other elements being zero. The process fidelity between the experimental result ($\chi_e$) and the identity process ($\chi_i$) was calculated by $F_p(\chi_e, \chi_i) = (Tr\{\sqrt{\sqrt{\chi_e}\chi_i\sqrt{\chi_e}}\})^2$.

**A.4 Measurements of CD, group index, and time-of-flight**

A free-space Mach–Zehnder interferometer, which was composed of a super-continuum source, two 50:50 plate beam splitters, two pairs of coupling lenses, two motorized stages, and an optical spectrum analyzer, was used to measure the group velocity dispersions and the group indices of the CTF and the SMF with short lengths (20 cm and 5 cm, respectively)[30]. The errors in measurement and polynomial fit were estimated and outlined in Fig. 4a.

We also built a single-photon time-of-flight measurement setup. The input photons were split into two routes, coupled into the two fibres with the same length (36.4 m), and then detected by two Si-SPADs. We used the heralding photons in the local Si-SPAD as the trigger to register the arrival time with a time stamp device (Swabian Instruments, with resolution of 10 ps). In order to eliminate the influence from electronic links, we carried out measurements twice with the two receiver Si-SPADs and cables being exchanged.

**A.5 Simulation of phase birefringence and PDL**

FEM simulations (COMSOL, Multiphysics) together with the fibre parameters extracted from the high-resolution SEM images yielded the effective modal indices and the confinement losses of the two fundamental modes of orthogonal polarizations. Their differences gave out the phase birefringence and the PDL.

**Acknowledgements:** This work is funded through the National Research and Development Program of China (2017YFB0405200, 2017YFA0303800), the National Natural Science Foundation of China (61575218, 61675011, 61827820, 61535009), Beijing Nova Program (Z181100006218097), and Research program of Beijing Municipal Education Commission (KZ201810005003).